\documentclass[default,iicol]{sn-jnl}

\usepackage{graphicx}%
\usepackage{multirow}%
\usepackage{amsmath,amssymb,amsfonts}%
\usepackage{amsthm}%
\usepackage{mathrsfs}%
\usepackage[title]{appendix}%
\usepackage{xcolor}%
\usepackage{textcomp}%
\usepackage{manyfoot}%
\usepackage{booktabs}%
\usepackage{algorithm}%
\usepackage{algorithmicx}%
\usepackage{algpseudocode}%
\usepackage{listings}%
\usepackage{natbib}

\theoremstyle{thmstyleone}%
\theoremstyle{thmstyletwo}%

\theoremstyle{thmstylethree}%

\begin{document}

\title[Article Title]{ImageBind-LLM: Multi-modality Instruction Tuning}


\author[1,2]{\fnm{Jiaming} \sur{Han}}\email{hanjiaming@pjlab.org.cn}
\equalcont{}

\author[1,2]{\fnm{Renrui} \sur{Zhang}}\email{zhangrenrui@pjlab.org.cn}
\equalcont{}

\author[1]{\fnm{Wenqi} \sur{Shao}}\email{shaowenqi@pjlab.org.cn}
\equalcont{}

\author[1‡]{\fnm{Peng} \sur{Gao}}\email{gaopeng@pjlab.org.cn}
\equalcont{}

\author[1]{\fnm{Peng} \sur{Xu}}\email{xupeng@pjlab.org.cn}
\equalcont{}

\author[1]{\fnm{Han} \sur{Xiao}}
\equalcont{Equal Contribution\quad $^{\ddagger}$Project Leader}

\author[1]{\fnm{Kaipeng} \sur{Zhang}}

\author[1]{\fnm{Chris} \sur{Liu}}

\author[1]{\fnm{Song} \sur{Wen}}

\author[1]{\fnm{Ziyu} \sur{Guo}}

\author[1,2]{\fnm{Xudong} \sur{Lu}}

\author[3]{\fnm{Shuai} \sur{Ren}}

\author[3]{\fnm{Yafei} \sur{Wen}}

\author[3]{\fnm{Xiaoxin} \sur{Chen}}

\author*[2]{\fnm{Xiangyu} \sur{Yue}}\email{xyyue@ie.cuhk.edu.hk}

\author*[2]{\fnm{Hongsheng} \sur{Li}}\email{hsli@ee.cuhk.edu.hk}

\author*[1]{\fnm{Yu} \sur{Qiao}}\email{qiaoyu@pjlab.org.cn}

\affil[1]{\orgname{Shanghai Artificial Intelligence Laboratory}, \orgaddress{\city{Shanghai}, \postcode{200030}, \country{China}}}

\affil[2]{\orgname{CUHK MMLab}, \orgaddress{\city{Hong Kong SAR}, \postcode{999077}, \country{China}}}

\affil[3]{\orgname{vivo AI Lab\vspace{-0.3cm}}, \orgaddress{\city{Shenzhen}, \postcode{518000}, \country{China}}}


\abstract{
We present \textbf{ImageBind-LLM}, a multi-modality instruction tuning method of large language models (LLMs) via ImageBind.
Existing works mainly focus on language and image instruction tuning, different from which, our ImageBind-LLM can respond to multi-modality conditions, including audio, 3D point clouds, video, and their embedding-space arithmetic by only image-text alignment training.
During training, we adopt a learnable bind network to align the embedding space between LLaMA and ImageBind's image encoder. Then, the image features transformed by the bind network are added to word tokens of all layers in LLaMA, which progressively injects visual instructions via an attention-free and zero-initialized gating mechanism. Aided by the joint embedding of ImageBind, the simple image-text training enables our model to exhibit superior multi-modality instruction-following capabilities. During inference, the multi-modality inputs are fed into the corresponding ImageBind encoders, and processed by a proposed visual cache model for further cross-modal embedding enhancement. The training-free cache model retrieves from three million image features extracted by ImageBind, which effectively mitigates the training-inference modality discrepancy.
Notably, with our approach, ImageBind-LLM can respond to instructions of diverse modalities and demonstrate significant language generation quality. Code is released at \url{https://github.com/OpenGVLab/LLaMA-Adapter}.
}




\maketitle

\begin{figure*}[t!]
  \centering
\includegraphics[width=\textwidth]{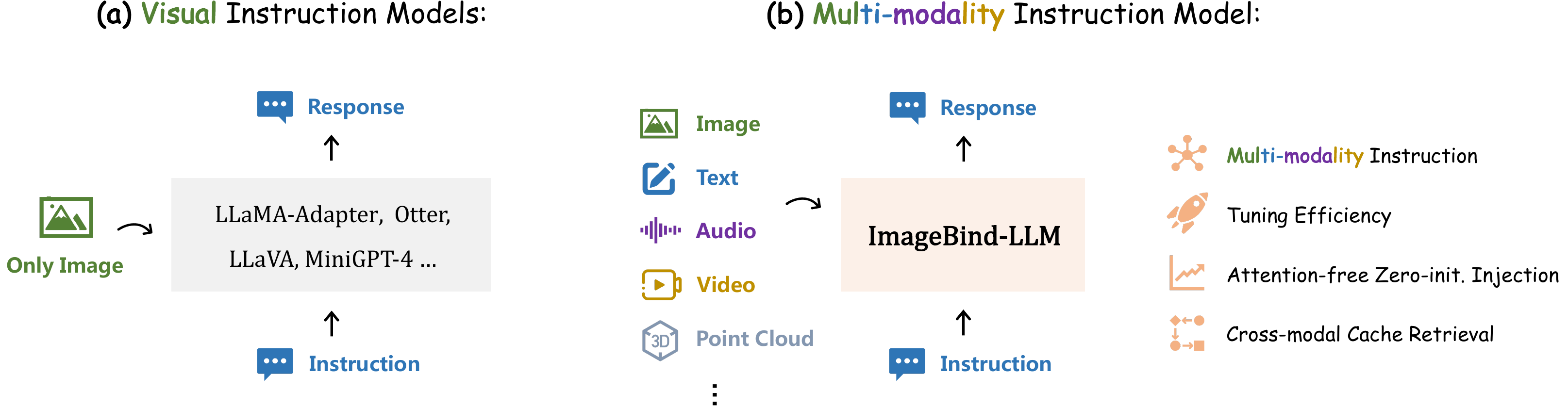}
   \caption{\textbf{Visual Instruction Models vs. Our Multi-modality ImageBind-LLM.} Different from existing works~\cite{gao2023llamaadapter,liu2023visual,zhu2023minigpt} conditioned only on image modality, ImageBind-LLM conducts a general multi-modality instruction tuning for image, text, audio, video, and 3D.}
    \label{fig1}
\end{figure*}

\section{Introduction}

Recently, we have witnessed substantial advancements in the instruction tuning of large language models (LLMs). With versatile intelligence and interactivity, ChatGPT~\cite{chatgpt} and GPT-4~\cite{OpenAI2023GPT4TR} present general-purpose chatting systems following human instructions in language and images, which is yet unreplicable due to the closed-source restriction. Inspired by this, Alpaca~\cite{alpaca}, LLaMA-Adapter~\cite{zhang2023llama}, and follow-up works~\cite{alpaca_lora,vicuna2023,peng2023instruction} propose to fine-tune the publicly available LLaMA~\cite{touvron2023llama} into language instruction models by self-constructed data. Further, to achieve image instruction tuning, LLaVA~\cite{liu2023visual}, LLaMA-Adapter~\cite{zhang2023llama}, and others~\cite{zhu2023minigpt} incorporate visual understanding capabilities into LLMs for image-conditioned generation. Despite the effectiveness of existing instruction tuning approaches, how to develop an LLM for general multi-modality instructions, e.g., text, image, audio, 3D point clouds, and video, is still under-explored. 

In this paper, we introduce a multi-modality instruction-following model, \textbf{ImageBind-LLM}, which efficiently fine-tunes LLaMA, guided by the joint embedding space in the pre-trained ImageBind~\cite{girdhar2023imagebind}. As compared in Figure~\ref{fig1}, different from previous visual instruction models (a), our ImageBind-LLM (b) can respond to input instructions of multiple modalities besides images, indicating promising extensibility and generalization capacity. 
Specifically, thanks to the image-aligned multi-modality embedding space of ImageBind, we propose to only leverage the vision-language data for multi-modality instruction tuning. For an image-caption pair, we first utilize the frozen image encoder of ImageBind to extract the global image feature, and adopt a learnable bind network for embedding transformation. Then, the transformed image feature is added to the word tokens at all transformer layers in LLaMA, which provides visual conditions to generate the corresponding textual caption.
Different from the zero-initialized attention in LLaMA-Adapter series~\cite{zhang2023llama,gao2023llamaadapter},
our visual injection method is attention-free and simply weighted by a trainable zero-initialized gating factor. In such an efficient manner, the instruction cues of ImageBind's multi-modality embeddings can be progressively injected into LLaMA as the training goes on, without disturbing the original language knowledge.

After the simple vision-language training, our ImageBind-LLM obtains the capability to follow instructions of various modalities, by applying ImageBind for modality-specific encoding, e.g., text, image, audio, and video. 
For instructions in 3D domains, we utilize the pre-trained 3D encoder in Point-Bind~\cite{guo2023point} to encode the input 3D point clouds.
To alleviate the modality discrepancy of image training and text/audio/3D/video-conditioned generation, we further propose a training-free visual cache model for embedding enhancement during inference. The cache model contains millions of image features in the training datasets extracted by ImageBind, which improves text/audio/3D/video embeddings by retrieving similar visual features, referring to Tip-Adapter~\cite{zhang2021tip}. This contributes to higher-quality language responses to multi-modality instructions. In diverse scenarios, we evaluate the multi-modality instruction-following capabilities of ImageBind-LLM, and observe consistent superior performance.

Overall, our ImageBind-LLM exhibits four main characteristics as follows.

\begin{itemize}
   \item \textbf{Multi-modality Instructions.} 
   Different from previous language and image instruction models, ImageBind-LLM is tuned to respond to general multi-modality inputs, such as image, text, audio, 3D point clouds, video, and their embedding-space arithmetic encoded by ImageBind and Point-Bind.
   
   \item \textbf{Tuning Efficiency.}
   During training, we freeze the image encoder of ImageBind, and fine-tune partial weights in LLaMA by parameter-efficient techniques, including LoRA~\cite{hu2021lora} and bias-norm tuning~\cite{xie2023difffit,zaken2021bitfit,frankle2020training,giannou2023expressive,gao2023llamaadapter}. Besides, we only train the additional bind network and zero-initialized gating factors.

   \item \textbf{Attention-free Zero-initialized Injection.}
   Instead of incorporating new instruction cues by attention layers, we directly add the multi-modality conditions with all word tokens of LLaMA, and adopt a learnable gating mechanism for progressive knowledge injection, more simple and effective.

   \item \textbf{Cross-modality Cache Retrieval.}
   To alleviate the modality discrepancy of training (only image) and inference (multiple modalities), we introduce a visual cache model constructed by ImageBind-extracted image features, which conducts cross-modality retrieval for embedding enhancement.

   
\end{itemize}

\section{Related Work}

\subsection{Visual Instruction Models.}
Given the rapid development of language instruction-following capabilities~\cite{touvron2023llama,alpaca,alpaca_lora}, how to enable large language models (LLMs) to perform visual understanding has also gained significant attention. LLaMA-Adapter~\cite{zhang2023llama}, for the first time, proposes to generate language responses conditioned on image inputs. It leverages a pre-trained encoder to extract image tokens, and incorporates them with LLaMA by parameter-efficient fine-tuning, which however can only tackle some naive visual question answering scenarios, i.e., ScienceQA~\cite{scienceqa}. For more general visual instruction-following circumstances, many efforts have been made to produce high-quality vision-language data for training by ChatGPT~\cite{chatgpt} or GPT-4~\cite{gpt4}, such as LLaVA~\cite{llava}, MiniGPT-4~\cite{zhu2023minigpt}, and Otter~\cite{li2023otter}. They normally follow the architecture of BLIP-2~\cite{li2023blip} with a more advanced Vicuna~\cite{vicuna2023}, or fine-tune the entire LLM with costly training resources. LLaMA-Adapter~\cite{zhang2023llama} develops a joint training strategy that only requires a combination of image-caption pairs and language instruction data, but still performs comparably to those with delicately constructed training data. VideoLLM~\cite{chen2023videollm} and Video-LLaMA~\cite{zhang2023video} also connect video reasoning modules with LLMs to allow for video instruction-following powers with temporal information. Different from them, our ImageBind-LLM takes a step forward by tuning a multi-modality LLM conditioned on language questions with image, video, audio, and 3D point cloud input, allowing for widespread applications.

\subsection{Multi-modality Alignment.}
Bridging different modalities within a joint embedding space for cross-modality processing has emerged as a critical research area in both vision and language. CLIP~\cite{radford2021learning}, ALIGN~\cite{jia2021scaling}, and Florence~\cite{yuan2021florence} utilize simple contrastive learning paradigms to align image and text pairs, contributing to promising zero-shot generalization performance. Flamingo~\cite{alayrac2022flamingo}, BLIP-2~\cite{li2023blip}, and MAGIC~\cite{zhang2022magic} adopt intermediate networks to connect pre-trained vision and language encoders. AudioCLIP~\cite{guzhov2022audioclip} and PointCLIP~\cite{zhang2022pointclip} respectively extend the embedding space of CLIP to other modalities, such as audio and 3D point clouds. Recently, ImageBind~\cite{girdhar2023imagebind} is proposed to share a single latent space with various modalities, including image, video, text, and audio. Inspired by ImageBind, Point-Bind~\cite{guo2023point} learns to blend 3D point cloud modalities into ImageBind, and achieves favorable 3D zero-shot accuracy. In this paper, we focus on aligning the shared embedding space in ImageBind/Point-Bind with LLaMA for multi-modality instruction-following capacity. PandaGPT~\cite{su2023pandagpt} also aims to tune a multi-modality LLM based on ImageBind, which cannot support 3D point clouds as input, and utilizes a stronger LLM, Vicuna~\cite{vicuna2023}, as the pre-trained language model. In contrast, our ImageBind-LLM is still based on LLaMA~\cite{touvron2023llama} and introduces unique attention-free zero-initialized injection with cross-modality cache retrieval for better multi-modality reasoning.

\section{Method}

In Section~\ref{s3.1}, we first briefly revisit some prior works as a preliminary, including ImageBind, cache models, and LLaMA-Adapter. Then, in Section~\ref{s3.2}, we introduce the details of our proposed multi-modality instruction tuning and cache-enhanced inference in ImageBind-LLM.

\subsection{A Revisit of Prior Works}
\label{s3.1}

\subsubsection{ImageBind}
With a single joint embedding space, ImageBind~\cite{girdhar2023imagebind} proposes to connect five different modalities, i.e., text, audio, depth, thermal, and Inertial Measurement Unit (IMU), all by image-paired data. Following CLIP~\cite{radford2021learning}, the pre-training of ImageBind adopts a contrastive loss, which clusters image features with other paired modalities, and pushes away unpaired ones in the embedding space. Self-supervised by large-scale image-paired data, ImageBind learns to encode different modalities into aligned feature embeddings, which obtains emergent cross-modal zero-shot capabilities. Then, ImageBind can be utilized to extend existing vision-language models to incorporate new modalities, such as text-to-audio/video retrieval, audio-to-image generation, and audio-referred object detection. Inspired by this image-centric property, our approach only conducts vision-language training to align the joint embedding space of ImageBind with LLaMA~\cite{touvron2023llama}, achieving efficient multi-modality instruction tuning.

\subsubsection{LLaMA-Adapter}
As a novel parameter-efficient fine-tuning method, LLaMA-Adapter~\cite{zhang2023llama} transforms LLaMA into a language instruction model by only 1.2M parameters within 1 hour, which exhibits comparable performance to the fully fine-tuned Alpaca~\cite{alpaca}. On top of this, LLaMA-Adapter~\cite{zhang2023llama} is also proposed to attain superior visual instruction-following capacity. It adopts a joint training paradigm for image-text and language-only instruction data, and still features tuning efficiency by updating partial parameters (14M) in LLaMA. One of the core innovations of LLaMA-Adapter series is the zero-initialized attention mechanism. 
They encode vision instruction signals as tokens, and concatenate them with the word tokens in LLaMA as prefixes.
Within every attention layer, a learnable gating factor is utilized to adaptively control how much information the new instruction knowledge is incorporated into LLMs.
Our ImageBind-LLM also adopts a zero-gated injection strategy for multi-modality instructions, but in a more simple and effective attention-free manner.

\subsubsection{Cache Models}
Without any training, a cache model can be utilized to store the features and labels of a training set, organizing them as a key-value database. During inference, the test sample serves as a query to retrieve from the keys and aggregate informative values via the key-query similarity. Starting from the conventional $k$ Nearest Neighbors algorithm ($k$-NN), cache models have been widely adopted to assist deep neural networks in language~\cite{khandelwal2019generalization}, 2D vision~\cite{vinyals2016matching}, and 3D point clouds~\cite{zhang2023parameter}.
Tip-Adapter~\cite{zhang2021tip} and its follow-up works~\cite{zhang2023parameter,udandarao2022sus,zhu2023not} propose to store the CLIP-extracted image features of the given few-shot data, and regard the cache model as a non-parametric adapter for downstream tasks. Similarly, we cache the ImageBind-extracted 1 million image features as both keys and values, which enhances the multi-modality embeddings in inference time.

\begin{figure*}[t!]
  \centering
\includegraphics[width=0.9\textwidth]{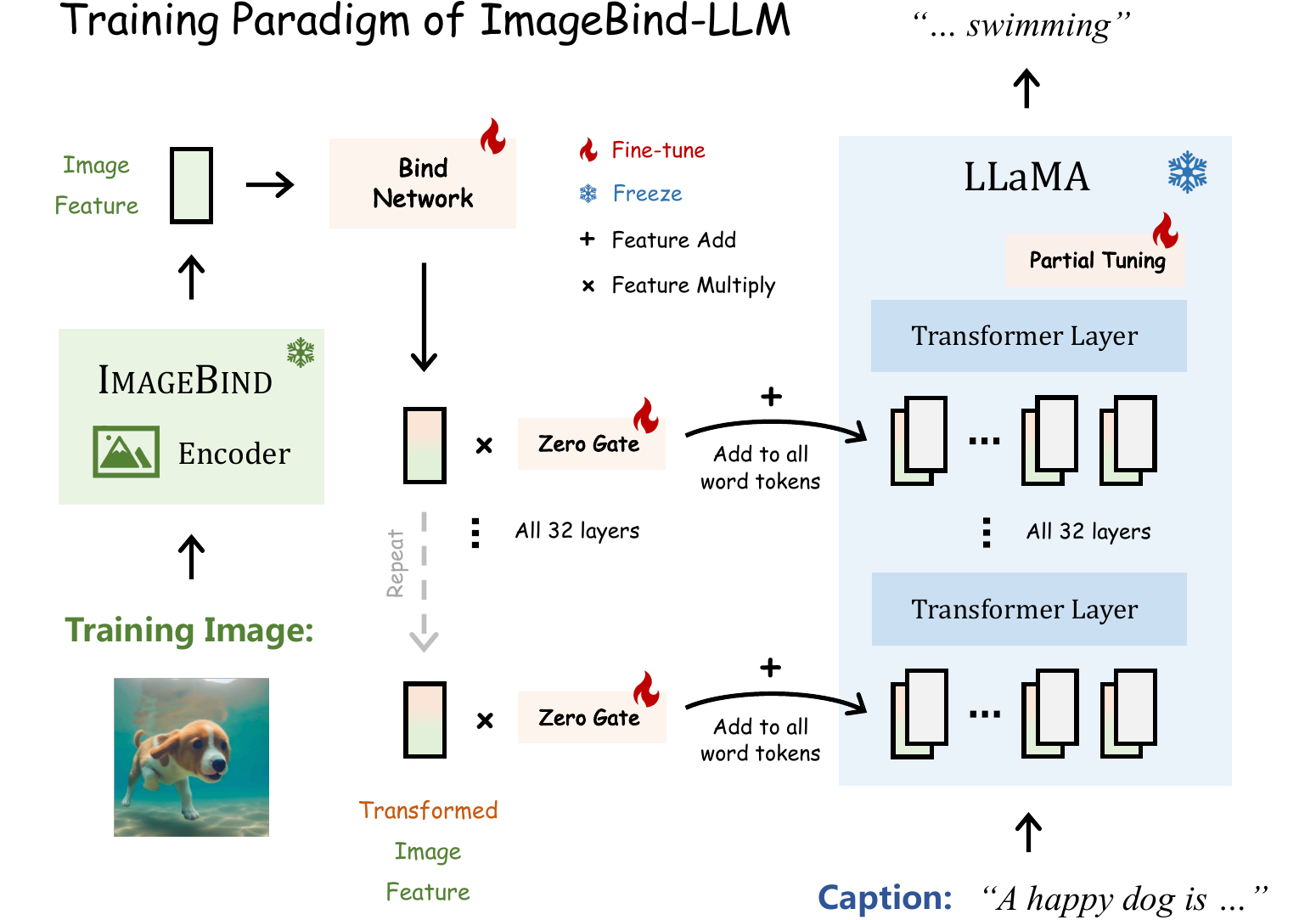}
   \caption{\textbf{Training Paradigm of ImageBind-LLM.} Considering the joint embedding space in imageBind~\cite{girdhar2023imagebind}, we only utilize image-text datasets for multi-modality instruction tuning of LLaMA~\cite{touvron2023llama}. We introduce a bind network for embedding alignment, and an attention-free zero-initialized mechanism for visual knowledge injection.}
   \vspace{0.2cm}
    \label{fig2}
\end{figure*}

\subsection{ImageBind-LLM}
\label{s3.2}

To obtain a multi-modality instruction model, we propose ImageBind-LLM, which includes two training stages: vision-language pre-training on image-caption data (Section~\ref{s3.2.1}) and multi-modality instruction tuning on visual instruction data (Section~\ref{subsubsec:instruction_tuning}). Besides, we also propose cross-modality cache retrieval for enhanced inference (Section~\ref{subsubsec:cache_inference}). The overall training paradigm of ImageBind-LLM is shown in Figure~\ref{fig2}.

\subsubsection{Vision-Language Pre-training}
\label{s3.2.1}

Given the modality-bound property of ImageBind~\cite{girdhar2023imagebind}, we only fine-tune LLaMA~\cite{touvron2023llama} to generate language responses conditioned on ImageBind-encoded images, after which, the model can inherently understand instructions of other modalities via the respective ImageBind encoders. Therefore, we propose to only leverage vision-language data for tuning a multi-modality instruction model. Following LLaMA-Adapter~\cite{zhang2023llama}, we adopt a two-stage training pipeline for ImageBind-LLM: first utilizing large-scale image-caption data~\cite{schuhmann2022laion,sharma2018conceptual,changpinyo2021conceptual} to learn the image-conditioned response capacity, then leveraging instruction-following data~\cite{llava,zhu2023minigpt} to preserve the long-sentence generation quality.
The overall training paradigm of ImageBind-LLM is shown in Figure~\ref{fig2}. For a given image-caption pair, we first adopt the frozen image encoder of ImageBind to extract the global visual feature. Then, we transform the visual feature with a learnable bind network, and add it to every word token in LLaMA. In an attention-free zero-initialized manner, LLaMA is injected by image condition and generates the given image caption.

\begin{figure*}[t!]
  \centering
\includegraphics[width=\textwidth]{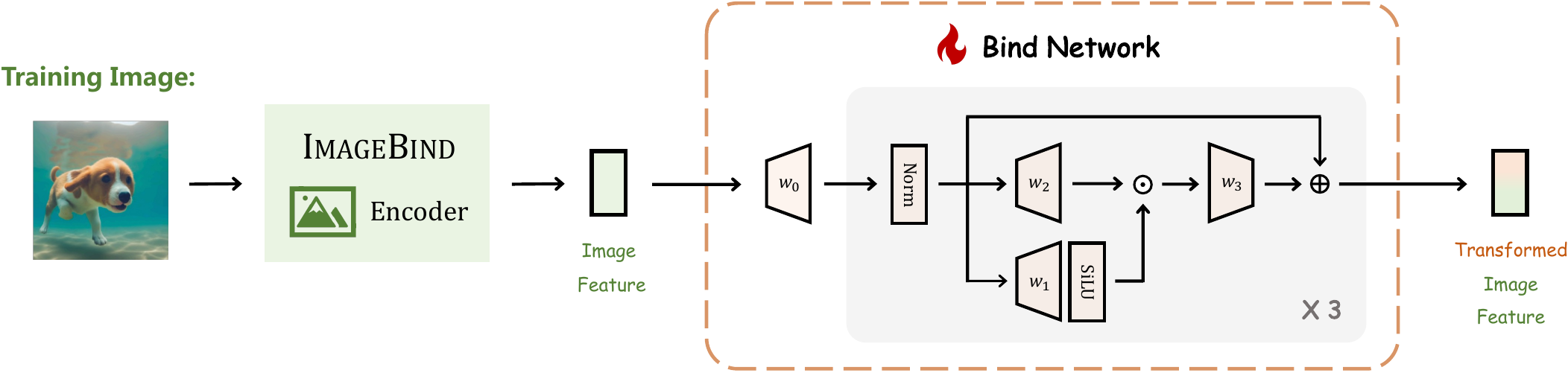}
   \caption{\textbf{Details of the Bind Network.} Referring to the Feed-Forward Network (FFN) in LLaMA~\cite{touvron2023llama}, we adopt cascaded blocks of RMSNorm~\cite{zhang2019rmsnorm}, SiLU activation functions~\cite{hendrycks2016gelu}, and residual connections~\cite{resnet}.
   This aims to align the image feature from ImageBind~\cite{girdhar2023imagebind} with LLaMA's word embeddings.}
    \label{fig3}
    \vspace{0.2cm}
\end{figure*}

\paragraph{Bind Network.}
In Figure~\ref{fig3}, we present the details of the bind network, which aims to align the embedding space between ImageBind and LLaMA. Specifically, we denote the $C_I$-dimensional global image feature encoded by ImageBind as $F_I \in \mathbb{R}^{1\times C_I}$. In the bind network, we first adopt a linear projection layer with a weight matrix $w_0 \in \mathbb{R}^{C_I\times C}$, formulated as $F_I^0 = F_I w_0 \in \mathbb{R}^{1\times C}$, where $C$ denotes the feature dimension of LLaMA. 
Inspired by the Feed-Forward Network (FFN) in LLaMA, we then cascade three projection blocks with RMSNorm~\cite{zhang2019rmsnorm}, SiLU activation functions~\cite{hendrycks2016gelu}, and residual connections~\cite{resnet}. For the $(i+1)$-th block with $F_I^{i}$ as input, we formulate the calculation of $F_I^{i+1}$ as (the normalization is omitted for simplicity)
\begin{align}
    F_I^{i+1} = F_I^{i} + (F_I^{i} w_2 \cdot \operatorname{SiLU}(F_I^{i} w_1)) w_3,\ \ \ 0 \le i < 3
\end{align}
where $w_1, w_2 \in \mathbb{R}^{C\times C_{h}}$ and $w_3 \in \mathbb{R}^{C_{h}\times C}$, with $C_{h}$ denoting the hidden dimension. After the bind network, we obtain the transformed image feature, $T_I \in \mathbb{R}^{1\times C}$, which learns to align the embedding space from ImageBind to LLaMA.

\paragraph{Attention-free Zero-initialized Injection.}
With the encoded image feature $T_I$, existing visual instruction methods, e.g., LLaMA-Adapter~\cite{zhang2023llama}, LLaVA~\cite{llava}, and MiniGPT-4~\cite{zhu2023minigpt}, concatenate it as the prefix to the word token sequence $\{T_W^j\}_{j=1}^N$ in LLaMA, where $N$ denotes the sequence length. Then, they leverage self-attention mechanisms in LLaMA's transformer blocks for visual knowledge incorporation from $T_I$ to $\{T_W^j\}_{j=1}^N$. However, such an attention-based approach not only causes extra computation budget, but also increases the training difficulty. In our ImageBind-LLM, we adopt a simpler and more effective method by attention-free zero-initialized injection. We directly add the image feature $T_I$ with every word token at all transformer layers of LLaMA, which explicitly fuses the visual conditions (and multi-modality inputs during inference) with the language knowledge in LLM. In addition, to adaptively control the level of integration, we utilize a learnable gating factor initialized by zero, denoted as $g_{zero}$.
For any word token $T_W^j$ in LLaMA, we formulate the visual injection as
\begin{align}
    T^j = T_I \cdot g_{zero} + T_W^j.
\end{align}
Similar to the zero-initialized attention in LLaMA-Adapter~\cite{zhang2023llama}, this gating factor can progressively increase during training, and inject more visual semantics into LLaMA, contributing to stable learning in the early training stage.

\begin{figure*}[t!]
  \centering
\includegraphics[width=\textwidth]{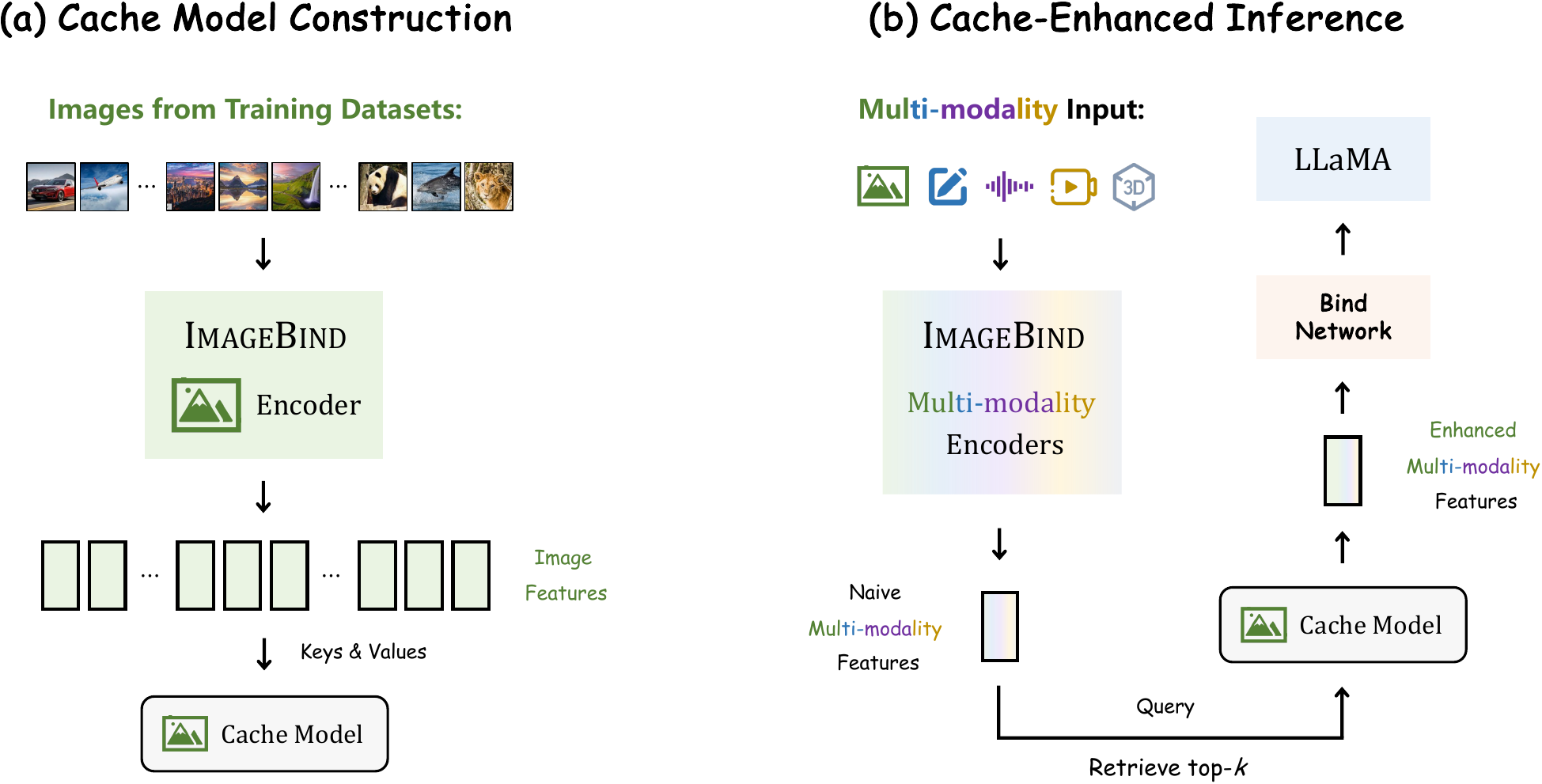}
   \caption{\textbf{Cache Retrieval for Inference.} To mitigate the training-inference discrepancy, we construct a training-free visual cached model of ImageBind-encoded image features (b). Then, during inference, we enhance the multi-modality embeddings by retrieving top-$k$ similar visual features in the cache model.}
    \label{fig4}
\end{figure*}

\subsubsection{Multi-modality Instruction Tuning}
\label{subsubsec:instruction_tuning}
Since we have connected ImageBind and LLaMA with a bind network via large-scale image-text pre-training, ImageBind-LLM can understand multimodal inputs (audio, video, and 3D point clouds), and generate language response conditioned on multi-modality inputs. However, unlike LLaVA~\cite{llava} and MiniGPT-4~\cite{zhu2023minigpt} that directly utilize a well-trained language instruction model Vicuna~\cite{vicuna2023} as the base LLM, we instead adopt a non-instruction model LLaMA. Therefore, in the second training stage, we partially tune the parameters in LLaMA to equip it with instruction-following ability, while keep the multi-modality encoders of ImageBind and the bind network frozen.

\paragraph{Parameter-efficient Fine-tuning.} 
Our second-stage training data is a mixture of language instruction data~\cite{shareGPT,peng2023instruction} and visual instruction data~\cite{llava}. As language instruction data contains no paired images, we input a fake image (filled with zero) as the placeholder during training. To maintain the multi-modality understanding capacity of the first-stage training, we only fine-tune LLaMA with parameter-efficient methods including Low-Rank Adaptation (LoRA)~\cite{hu2021lora} and bias-norm tuning~\cite{xie2023difffit,zaken2021bitfit,frankle2020training,giannou2023expressive,gao2023llamaadapter}. Specifically, we add a low-rank layer for each linear layer in the transformer, where the rank is set to 16 by default. We also unfreeze all the normalization layers and add a learnable bias term to the linear layers. The parameters of all other modules are frozen during training.

\paragraph{High-Quality Instruction Tuning.} 
Although the fine-tuned ImageBind-LLM can generate instruction-following responses, we notice that it occasionally fantasizes about objects that don't exist in the input modality. Therefore, we introduce additional instruction tuning stage using high-quality instruction data from MiniGPT-4~\cite{zhu2023minigpt}. Different from the visual instruction data generated by ChatGPT/GPT4, MiniGPT-4 manually collects 3.5K image description data for high-quality vision-language alignment. Here we also adopt the 3.5K image description data for further instruction tuning, which only takes a few minutes.

\subsubsection{Cache Retrieval for Inference}
\label{subsubsec:cache_inference}

After fine-tuning by visual instruction data, ImageBind-LLM can inherently derive the multi-modality instruction-following capacity. Then, besides the naive inference mode (Figure~\ref{fig4} (a)), we further propose to construct a visual cache model by ImageBind for multi-modality embedding enhancement (Figure~\ref{fig4} (b)).

\paragraph{Naive Multi-modality Inference.}
Via the bind network, the image features from ImageBind can be well aligned with LLaMA's word tokens. Given the joint multi-modality embedding space of ImageBind, our ImageBind-LLM spontaneously obtains the understanding capability for input instructions with various modalities, such as image, text, audio, and video, only if we apply their corresponding encoders from ImageBind before the bind network. 
For 3D point clouds, we can also utilize the pre-trained 3D encoder of Point-Bind~\cite{guo2023point} for global feature extraction, which shares the same embedding space with ImageBind.

\paragraph{Cache-enhanced Inference.}
Despite the effectiveness of the aforementioned naive mode, there exists modality discrepancy in ImageBind-LLM between training and inference. Namely, we adopt image encoder of ImageBind for training, but switch to other encoders for inference, which slightly disturbs the tuned bind network and LLaMA. Therefore, we construct a training-free cache model of image features to enhance the multi-modality embeddings during inference.
As shown in Figure~\ref{fig4} (a), we utilize ImageBind to encode a subset of the vision-language training data, and store them as both keys and values in the cache model. For an input multi-modality instruction in Figure~\ref{fig4} (b), we regard its ImageBind-encoded feature as the query, $F_M \in \mathbb{R}^{1\times C_I}$, and retrieve the top-$k$ similar visual keys from the cache model, denoted as $F_{key} \in \mathbb{R}^{k\times C_I}$.
We formulate the top-$k$ cosine similarity as
\begin{align}
    S_{topk} = F_M F_{key}^T\ \ \in \mathbb{R}^{1\times k},
\end{align}
where we suppose $F_M$ and $F_{key}$ have been L2-normalized. Then, according to $S_{topk}$, we aggregate the corresponding cached values, $F_{value} \in \mathbb{R}^{k\times C_I}$ (top-$k$ similar image features), and add the result to the original feature $F_M$ via a residual connection, formulated as
\begin{align}
    F_M^e = \alpha \cdot S_{topk} F_{value} + (1-\alpha) \cdot F_M,
\end{align}
where $\alpha$ serves as a balance factor. Aided by the cache model, the enhanced feature $F_M^e$ is adaptively incorporated with similar visual semantics from the cache model. This boosts the representation quality of other modalities, and mitigates their semantic gap to the images used for training. After this, $F_M^e$ is fed into the bind network for feature transformation and LLaMA for response generation.

\subsection{Advanced Applications}
Besides the superior multi-modality instruction-following capabilities, our ImageBind-LLM can also be extended to a wide range of advanced applications with simple modifications.

\subsubsection{Bilingual Instruction Tuning}
In addition to English instructions, ImageBind-LLM can be easily upgraded to a bilingual instruction-following model, e.g., English and Chinese. More specifically, we replace the basic LLM from LLaMA to a bilingual LLM, ChineseLLaMA\footnote{\url{https://github.com/OpenLMLab/OpenChineseLLaMA}} and add 52K Chinese instruction data from GPT4LLM~\cite{peng2023instruction} for joint instruction tuning. Although we do not have direct Chinese visual instruction data for the first vision-language training stage, we observe that our bilingual ImageBind-LLM implicitly learns the alignment between Chinese, English and multi-modality inputs, and can well follow Chinese instructions conditioned on other modality inputs.

\subsubsection{Any-to-any Generation}
Currently, most multi-modality instruction models are limited to generating only textual responses, lacking the ability to respond with other modal outputs, e.g., image, audio, and point clouds. Since ImageBind is an extension of CLIP~\cite{radford2021learning},
we can append CLIP-conditioned generative models after ImageBind's encoders, such as Stable Diffusion~\cite{rombach2022high}, Make-An-Audio~\cite{huang2023make}, and CLIP-Forge~\cite{sanghi2022clip}, respectively for image, audio, and point cloud generation. Instead of directly inputting ImageBind features into these generative models, we adopt cache-enhanced generation to mitigate the modality discrepancy, similar to the approach in Cache-enhanced Inference (Section~\ref{subsubsec:cache_inference}). In this way, we can achieve instruction models with any-to-any generation system, i.e., responding to multi-modality instructions by multi-modality responses.
as an example, our ImageBind-LLM can generate both textual and image responses for multi-modality inputs (\emph{e.g.}, image, audio and point clouds).

\subsubsection{Integration with Object Detection}
Visual instruction models can answer questions based on the global content of input images. However, they cannot associate the text response with regional objects in the image, which is important for fine-grained tasks such as visual reasoning and grounding. We provide a solution to connect ImageBind-LLM with object detectors~\cite{groundingdino}. For a response generated by ImageBind-LLM, we use traditional noun parsers~\cite{bird2009natural} or ChatGPT~\cite{chatgpt} to extract nouns in the response. Then we feed the input image and parsed nouns into object detectors to get object detection results. Generally, the traditional noun parser is enough for parsing meaningful nouns, but it cannot handle nouns with complex modifiers, such as "a running black dog". Therefore, we will also ask ChatGPT to extract complex nouns in the response. 

\subsubsection{ImageBind-LLM as Chatbot}
\label{subsubsec:chatbot}
ImageBind-LLM was originally designed as a single-turn multi-modality instruction model. We turn ImageBind-LLM into a multi-turn chatbot by training it on multi-turn conversation data, including language conversation data from ShareGPT~\cite{shareGPT} and visual conversation data from LLaVA~\cite{liu2023visual}. 
By this, ImageBind-LLM can be used as a multi-turn chat model to answer open-ended questions on multi-modality inputs.

\subsubsection{ImageBind-LLM for API Control}
In addition to its primary multimodal instruction-following capacity, ImageBind-LLM also exhibits the potential to invoke diverse API controls for multi-modality tool usage. To achieve this, we leverage the tool-related instruction dataset introduced in GPT4Tools~\cite{yang2023gpt4tools} to empower ImageBind-LLM with the ability to effectively utilize various tools. By training ImageBind-LLM on the GPT4Tools dataset using our proposed training paradigm, we observe its impressive proficiency in calling different APIs, enabling it to accomplish a wide range of tasks, even when encountering previously unseen tools. This performance in API control highlights the potential of ImageBind-LLM as a versatile visual assistant capable of solving diverse real-world problems.

\section{Experiment}
\subsection{Training Details}

\subsubsection{Datasets}
We train ImageBind-LLM on a collection of open-sourced image-text pair data, language-only and visual instruction data.

\noindent \textbf{Image-Text Pair Data.}
Our ImageBind-LLM is pre-trained on the concatenation of open-sourced image-text pair data, including COCO~\cite{chen2015cococap}, CC3M~\cite{sharma2018conceptual}, CC12M~\cite{changpinyo2021conceptual}, SBU~\cite{sbu}, LAION-2B~\cite{schuhmann2022laion}, COYO~\cite{kakaobrain2022coyo700m} and MMC4~\cite{zhu2023mmc4}.
Note that MMC4-Core~\cite{zhu2023mmc4} is a billion-scale corpus of images interleaved with text. We extract 20M high-quality image-text pairs from MMC4-Core according to the provided clip alignment score. For LAION-2B~\cite{schuhmann2022laion} dataset, we also extract 100M high-quality image-text pairs based on their CLIP alignment scores. The concatenation of all open-sourced image-text pairs result into 940M image-text pair data. Unlike BLIP~\cite{li2022blip} which designs an effective data cleaning pipeline, our image-text pairs are much noisy. However, we empirically observe strong image understanding and factual ability of ImageBind-LLM when pre-trained with this dataset. In the future, we will explore advanced approaches for data cleaning and deduplication. 


\noindent \textbf{Instruction Tuning Datasets.}
Our instruction tuning data includes language instruction data Alpaca~\cite{alpaca}, GPT4LLM~\cite{peng2023instruction} and ShareGPT~\cite{shareGPT}, visual instruction data LLaVA~\cite{llava} and MiniGPT4~\cite{zhu2023minigpt}. For language instruction data, Alpaca contains 52K single-turn instruction data collected from GPT3.5; GPT4LLM is a GPT4 version of Alpaca with higher quality; ShareGPT is a collection of user-shared conversations with ChatGPT/GPT4. For visual instruction data, LLaVA adopts GPT4 to transform image captions or object detection annotations into 150K visual instruction data; MiniGPT4 curates a high-quality image description dataset with 3.5K examples. Note that we will convert multi-round conversation data into single turn data for instruction tuning.

\subsubsection{Implementation Details}
For cache-enhanced inference, we use the FAISS library~\cite{johnson2019faiss} to build our retrieval system, and the Autofaiss library\footnote{\url{https://github.com/criteo/autofaiss}} to find the optimal hyper-parameters for the index. By default, all images from CC3M~\cite{sharma2018conceptual} is used to build the cache model. We pre-train the model on 32 A100 GPUs for 3 epochs. The total batch size and learning rate is set to 1024 and 4e-4, respectively. We fine-tune the model on 8 A100 GPUs for 4 epochs The warmup epoch, total batch size, learning rate is set to 1, 32 and 1.25e-4.

\begin{table*}[t]
\centering
\small
\caption{\textbf{Zero-shot Performance on OCR, KIE, and Image Captioning Tasks.} Evaluation metrics include word accuracy for OCR datasets, entity-level F1 score for KIE datasets, and CIDEr score for image captioning datasets. ImageBind-LLM-D: ImageBind-LLM trained on multi-turn conversation data (Sec.~\ref{subsubsec:chatbot}).}
\label{tab:vka_results}
\resizebox{\textwidth}{!}{%
\begin{tabular}{c|c|ccccccc}
\toprule
\multicolumn{2}{c|}{Model} & BLIP2 & InstructBLIP & LA & LLaVA & PandaGPT & ImageBind-LLM & ImageBind-LLM-D \\
\midrule
\multicolumn{2}{c|}{\#Token}& 32   & 32           & 10    & 257   & 1        & 1             & 1               \\
\cmidrule(lr){1-2}\cmidrule(lr){3-9}
\multirow{12}{*}{OCR}
& IIIT5K      & 80.17 & 83.90 & 36.30 & 31.57 & 5.27 & 13.9  & 13.87 \\
& IC13        & 81.13 & 82.08 & 20.87 & 16.39 & 4.60 &  7.43 &  7.19 \\
& IC15        & 66.68 & 73.57 & 29.40 & 26.58 & 4.57 & 11.94 & 11.36 \\
& Total-Text  & 68.31 & 71.51 & 30.93 & 24.51 & 4.06 & 10.79 & 10.11 \\
& CUTE80      & 85.07 & 86.11 & 35.76 & 36.46 & 6.60 & 20.14 & 20.83 \\
& SVT         & 85.78 & 86.86 & 20.40 & 18.55 & 3.40 &  8.35 &  7.11 \\
& SVTP        & 77.34 & 80.93 & 31.01 & 27.44 & 4.96 & 10.39 & 10.08 \\
& COCO-Text   & 53.62 & 58.25 & 20.94 & 18.05 & 2.67 &  5.59 &  5.12 \\
& WordArt     & 73.66 & 75.12 & 38.98 & 35.87 & 7.81 & 21.24 & 20.58 \\
& CTW         & 67.43 & 68.58 & 18.13 & 16.73 & 2.74 &  7.12 &  7.38 \\
& HOST        & 57.28 & 61.22 & 16.60 & 15.94 & 3.97 &  7.53 &  7.82 \\
& WOST        & 68.83 & 73.26 & 21.73 & 20.49 & 4.01 &  8.73 &  8.57 \\
\cmidrule(lr){1-2}\cmidrule(lr){3-9}
\multirow{2}{*}{KIE}
& SROIE & 0.08 & 0.09 & 0.02 & 0.01 & 0.01 & 0.01 & 0.01  \\
& FUNSD & 1.02 & 1.03 & 2.16 & 1.93 & 2.06 & 2.00 & 2.01  \\
\cmidrule(lr){1-2}\cmidrule(lr){3-9}
\multirow{2}{*}{Caption}
& NoCaps     & 48.58 & 46.33 & 41.66 & 33.09 & 29.65 & 30.43 & 29.64  \\
& Flickr-30k & 46.48 & 50.45 & 30.49 & 27.65 & 23.02 & 23.04 & 23.49  \\
\bottomrule
\end{tabular}%
}
\end{table*}


\begin{table*}[t]
    \centering
    \small
    \caption{\textbf{Zero-shot Performance on VQA, KGID, and VE Tasks.} For VQA and KGID tasks, Mean Reciprocal Rank (MRR) is used for the Visdial, while top-1 accuracy is employed for the remaining tasks.}
    \label{tab:vr_results}
    \resizebox{\textwidth}{!}{%
    \begin{tabular}{c|c|ccccccc}
    \toprule
    \multicolumn{2}{c|}{Model} & BLIP2 & InstructBLIP & LA & LLaVA & PandaGPT & ImageBind-LLM & ImageBind-LLM-D  \\
    \midrule
    \multicolumn{2}{c|}{\#Token}& 32   & 32           & 10    & 257   & 1        & 1             & 1               \\
    \cmidrule(lr){1-2}\cmidrule(lr){3-9}
    \multirow{9}{*}{VQA}
    & DocVQA  &  4.75 &  5.89 &  8.13 &  6.26 &  3.42 &  4.04 &  4.08  \\
    & TextVQA & 31.98 & 39.60 & 43.76 & 38.92 & 16.42 & 23.98 & 23.98  \\
    & STVQA   & 20.98 & 28.30 & 32.33 & 28.40 & 11.23 & 15.55 & 14.75  \\
    & OCR-VQA & 38.85 & 60.20 & 38.12 & 23.40 & 22.39 & 23.24 & 22.31  \\
    & OKVQA   & 44.93 & 60.52 & 55.93 & 54.36 & 50.85 & 51.66 & 51.70  \\
    & GQA     & 45.53 & 49.96 & 43.93 & 41.30 & 41.56 & 41.23 & 41.12  \\
    & Visdial & 10.73 & 45.20 & 12.92 & 14.66 & 90.80 & 12.66 & 12.91  \\
    & IconQA  & 62.82 & 56.25 & 41.83 & 42.95 & 46.04 & 37.97 & 41.81  \\
    & VSR     & 63.63 & 41.28 & 50.63 & 51.24 & 46.75 & 49.37 & 49.78  \\
    \cmidrule(lr){1-2}\cmidrule(lr){3-9}
    \multirow{2}{*}{KGID}
    & ScienceQA IMG & 60.73 & 46.26 & 54.19 & 49.33 & 52.80 & 55.83 & 51.41  \\
    & VizWiz        & 65.44 & 65.31 & 62.07 & 62.42 & 46.95 & 51.90 & 51.28  \\
    \bottomrule
    \end{tabular}%
    }
    \end{table*}

\begin{table*}[t]
    \centering
    \small
    \caption{\textbf{Perception Performance Comparison on MME~\cite{fu2023mme} benchmark.} The full score for the overall perception tasks is 2000, while for the 10 subtasks is 200.}
    \label{tab:mme_results1}
    \setlength{\tabcolsep}{8pt}
    \begin{tabular}{l|cccccc}
    \toprule
    Model & MiniGPT-4 & Otter & LLaMA-Adapter & LLaVA &PandaGPT &ImageBind-LLM \\
    \midrule
    \#Token& 32   & 64           & 10    & 257   & 1        & 1\\
    \cmidrule(lr){1-1}\cmidrule(lr){2-7}
    Existence  &  115.00 & 48.33 &  120.00 &  50.00 &  70.00 &  128.33 \\
    Count & 123.33 & 50.00 & 50.00 & 50.00 & 50.00 & 60.00 \\
    Position   & 81.67 & 50.00 & 48.33 & 50.00 & 50.00 & 46.67 \\
    Color & 110.00 & 55.00 & 75.00 & 55.00 & 50.00 & 73.33 \\
    Poster   & 55.78 & 44.90 & 99.66 & 50.00 & 76.53 & 64.97 \\
    Celerity     & 65.29 & 50.00 & 86.18 & 48.82& 57.06 & 76.47 \\
    Scene & 95.75 & 44.25 & 148.50 & 50.00 & 118.00 & 113.25 \\
    Landmark  & 69.00 & 49.50 & 150.25 & 50.00 & 69.75 & 62.00 \\
    Artwork     & 55.75 & 41.75 &69.75 & 49.00 & 51.25 & 70.75 \\
    OCR     & 95.00 & 50.00 & 125.00 &50.00 & 50.00 & 80.00 \\
    \cmidrule(lr){1-7}
    Perception & 866.58 & 483.73 & 972.67 & 502.82 & 642.59 & 775.77 \\
    \bottomrule
    \end{tabular}%
    \end{table*}

\begin{table*}[t]
    \centering
    \small
    \caption{\textbf{Cognition Performance Comparison on MME~\cite{fu2023mme} benchmark.} The full score for the overall perception tasks is 800, while for the 4 subtasks is 200.}
    \label{tab:mme_results2}
    \resizebox{\textwidth}{!}{%
    \begin{tabular}{l|cccccc}
    \toprule
    Model & MiniGPT-4 & Otter & LLaMA-Adapter & LLaVA &PandaGPT &ImageBind-LLM \\
    \midrule
    \#Token& 32   & 64           & 10    & 257   & 1        & 1\\
    \cmidrule(lr){1-1}\cmidrule(lr){2-7}
    Commonsense Reasoning &  72.14 &  38.57 & 81.43 & 57.14 &73.57 &48.57 \\
    Numerical Calculation & 55.00 &20.00 &62.50 &50.00 &50.00 & 55.00 \\
    Text Translation   & 55.00 & 27.50 & 50.00 & 57.50 & 57.50 & 50.00 \\
    Code Reasoning & 110.00 & 50.00 & 55.00 & 50.00 & 47.50 & 60.00 \\
    \cmidrule(lr){1-7}
    Cognition & 292.14 & 136.07 & 248.93 &214.64 & 228.57 &213.57 \\
    \bottomrule
    \end{tabular}%
    }
    \end{table*}

\subsection{Quantitative Evaluation on Traditional Tasks}
In this section, we conducted quantitative evaluations of ImageBind-LLM on 27 datasets using a zero-shot approach. Our quantitative evaluation encompassed five specific tasks: Optical Character Recognition (OCR), Key Information Extraction (KIE), Image Captioning, Visual Question Answering (VQA), and Knowledge-Grounded Image Description (KGID). Notably, all these tasks are evaluated following a VQA-style approach. The comparisons of ImageBind-LLM with other well-known Vision-Language Models (VLMs) such as BLIP2 \cite{li2023blip}, InstructBLIP \cite{dai2023instructblip}, LLaVA \cite{llava}, LLaMA-Adapter (LA)~\cite{zhang2023llama}, and multi-modality LLM model PandaGPT \cite{su2023pandagpt} are presented in Table \ref{tab:vka_results} and Table \ref{tab:vr_results}. 

\subsubsection{Experimental Settings}

\noindent \textbf{OCR Tasks.}
We evaluate ImageBind-LLM on 12 representative OCR datasets, including IIIT5K~\cite{ittt5k}, ICDAR 2013(IC13)~\cite{ic13}, ICDAR 2015 (IC15)~\cite{ic15}, Total-Text~\cite{total-text}, CUTE80~\cite{cute80}, Street View Text (SVT)~\cite{svt}, SVTP-Perspective (SVTP)~\cite{svtp}, COCO-Text~\cite{coco-text}, WordArt~\cite{wordart}, SCUT-CTW1500 (CTW)~\cite{ctw}, Heavily Occluded Scene Text (HOST)~\cite{ost}, Weakly Occluded Scene Text (WOST)~\cite{ost}. These datasets encompass a diverse collection of images containing textual information, enabling a comprehensive comparison between models. The evaluation of model performance was based on top-1 accuracy, using the prompt "What is written in the image?"

\noindent \textbf{KIE Tasks.}
We evaluate ImageBind-LLM on 2 KIE benchmarks, including SROIE~\cite{sroie} and FUNSD~cite{funsd}. These benchmarks encompass a diverse range of document types, including receipts and forms, which necessitate the extraction of specific information. The evaluation of models involved using entity-level F1 scores. To further enhance the evaluation process, we employed prompts tailored to the specific information that the model was required to extract. For instance, in the case of the SROIE benchmark, prompts such as "What is the name of the company that issued this invoice?" were used to extract company information, while prompts like "Where was this invoice issued?" were employed to extract address information.

\noindent \textbf{VQA Tasks.} 
We employ 9 benchmarks in the VQA task, namely DocVQA~\cite{docvqa}, TextVQA~\cite{textvqa}, STVQA~\cite{stvqa}, OCR-VQA~\cite{ocr-vqa}, OKVQA~\cite{okvqa}, GQA~\cite{gqa}, IconQA~\cite{iconqa}, Visual Spatial Reasoning (VSR)~\cite{vsr}, and Visual Dialog (Visdial)~\cite{visdial}. These benchmarks encompass a diverse collection of question-image pairs that cover a wide range of topics. The task requires models not only to comprehend the visual content but also to understand and reason about the questions presented. For specific evaluation purposes, we utilize the Mean Reciprocal Rank (MRR) metric for Visdial and top-1 accuracy for the remaining datasets. These metrics provide valuable insights into the model's proficiency in accurately answering questions across the various VQA benchmarks.

\noindent \textbf{KGID tasks.} 
The KGID task aims to assess the model's ability to produce descriptive and precise image captions by incorporating external knowledge. To evaluate performance in this task, we utilize the ScienceQA~\cite{scienceqa} and VizWiz~\cite{vizwiz} benchmarks, which include images accompanied by textual descriptions and knowledge-based information. It is worth mentioning that, for ScienceQA, we specifically consider only those samples that contain images.

\subsubsection{Analysis}
Table \ref{tab:vka_results} and Table \ref{tab:vr_results} clearly demonstrate the exceptional zero-shot performance of ImageBind-LLM across all evaluated tasks. When it comes to OCR, Image Captioning, and KGID, ImageBind-LLM achieved competitive performance compared with other VLMs and outperformed PandaGPT, thus showcasing the effectiveness of ImageBind-LLM's modality alignment strategy. Furthermore, ImageBind-LLM also delivered an impressive performance on KIE and VQA datasets.

Further investigating the reason behind ImageBind-LLM's relatively better performance than PandaGPT, we delve into the implementation details of ImageBind-LLM and PandaGPT. Firstly, we observe a significant disparity in ImageBind-LLM and PandaGPT's utilization of the ImageBind extracted feature. PandaGPT employs a single linear projection layer for processing the ImageBind extracted feature, whereas ImageBind-LLM employs a bind network, which potentially facilitates better alignment between language and modalities through ImageBind. Another distinction lies in their choice of LLM model, with PandaGPT utilizing Vicuna and ImageBind-LLM employing LLaMA. Notably, Vicuna, being tuned based on LLaMA and possessing a higher Elo rating as indicated in \cite{vicuna2023}, potentially enhances PandaGPT's language comprehension and response generation capabilities.

Then for why both ImageBind-LLM and PandaGPT have a poor OCR ability compared to other VLMs, we discovered that both of them employ only one token for the modality feature, while the other VLMs utilize at least ten tokens for capturing visual information. This disparity may allow other VLM models to better comprehend the visual information depicted in the images.

These results not only highlight the remarkable zero-shot performance of ImageBind-LLM in various vision and language tasks but also underscore its ability to comprehend and generate accurate responses in diverse scenarios. Moreover, the model's adeptness in multi-modality understanding further demonstrates its potential as a robust and versatile solution for real-world applications.

\subsection{Quantitative Evaluation on MME Benchmark}

\subsubsection{Experimental Settings}
In contrast to traditional multi-modality tasks, we also evaluate our ImageBind-LLM on a newly proposed benchmark, MME~\cite{fu2023mme}, which is specially deigned for the recent VLMs. MME benchmark systematically measures two multi-modality capabilities of existing methods: perception and cognition. The former with 10 subtasks refers to recognizing specific objects in images, while the latter with 4 subtasks is more challenging for deducing complex answers from visual information. For each test image, MME adopts an instruction of a question and a description ``Please answer yes or no'', which prompts LLMs to answer ``yes'' or ``no''. Such a concise instruction-answer evaluation allows for fair comparison of LLMs without the impact of prompt engineering.

\subsubsection{Analysis}
In Table~\ref{tab:mme_results1} and~\ref{tab:mme_results2}, we respectively show the performance comparison of different VLMs on MME's perception and cognition tasks, including MiniGPT-4~\cite{zhu2023minigpt}, Otter~\cite{li2023otter}, LLaMA-Adapter~\cite{zhang2023llama}, LLaVA~\cite{llava}, and PanadaGPT~\cite{su2023pandagpt}. As shown, MiniGPT-4 can achieve the best scores since it is trained upon a pre-trained BLIP-2~\cite{li2023blip}. Otter and PandaGPT are developed based on OpenFlamingo~\cite{openflamingo} and Vicuna~\cite{vicuna2023}, which endow them with well-initialized language processing abilities. Instead, similar to LLaMA-Adapter, our ImageBind-LLM is fine-tuned on the original LLaMA model, and still performs competitively to others. Especially on `Existence' and `Artwork', ImageBind-LLM outperforms the second-best methods by +8.33 and +1.00 scores, respectively. Overall, our approach is more expert at the `Perception' tasks, ranking the third place and surpassing another multi-modality model, PandaGPT, by +133.18 score. As analyzed above in Section~\ref{s3.2}, we believe our performance can be further improved if using more multi-modality tokens fed into LLMs.

\begin{figure*}[htbp]
    \includegraphics[width=\textwidth]{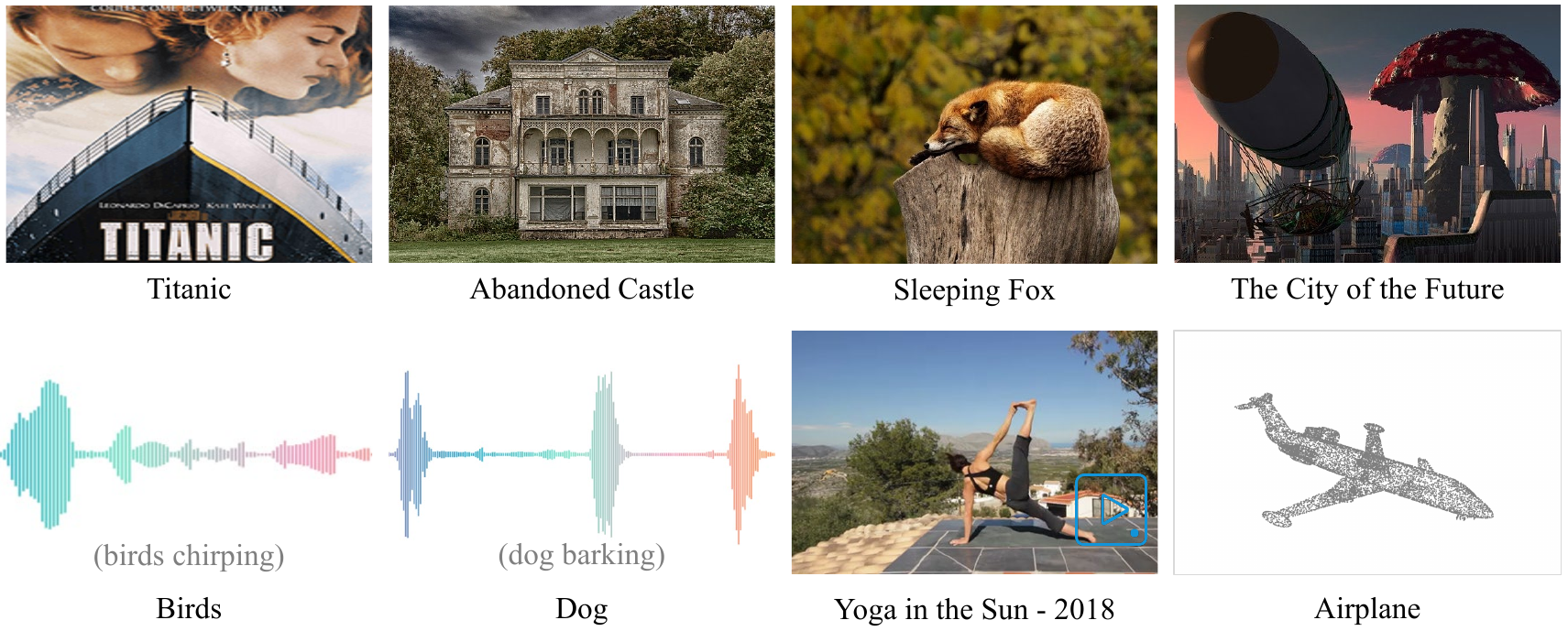}
    \caption{\textbf{Multi-modal Captioning Results} with ImageBind-LLM. The training data is a collection of image-text pairs. We only train the parameters of the bind network during this stage.}
    \label{fig:image_caption}
\end{figure*}

\begin{figure*}[htbp]
    \includegraphics[width=\textwidth]{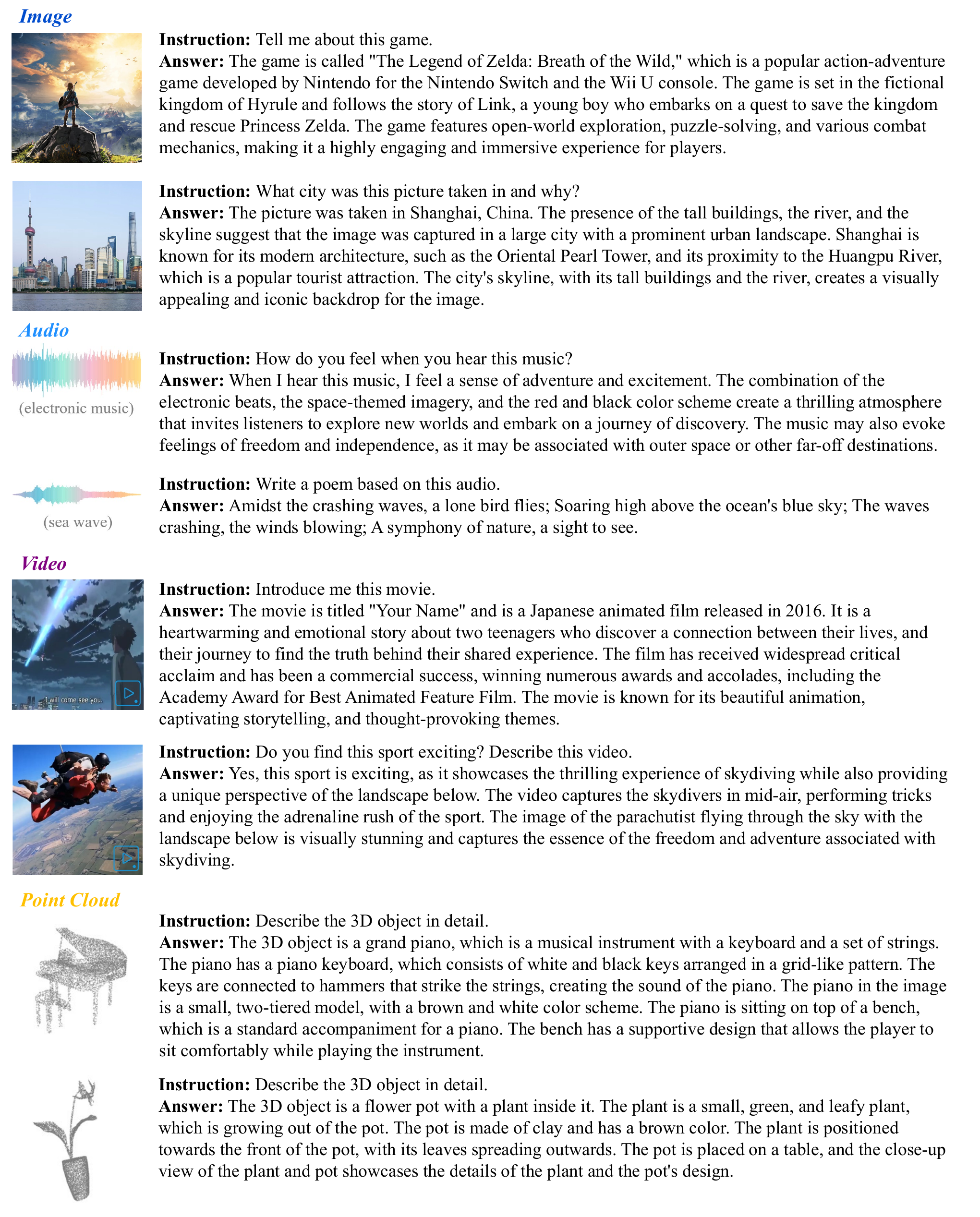}
    \caption{\textbf{Multi-modal Instruction Examples with ImageBind-LLM}.}
    \label{fig:instruction_demo}
\end{figure*}

\begin{figure*}[htbp]
    \includegraphics[width=\textwidth]{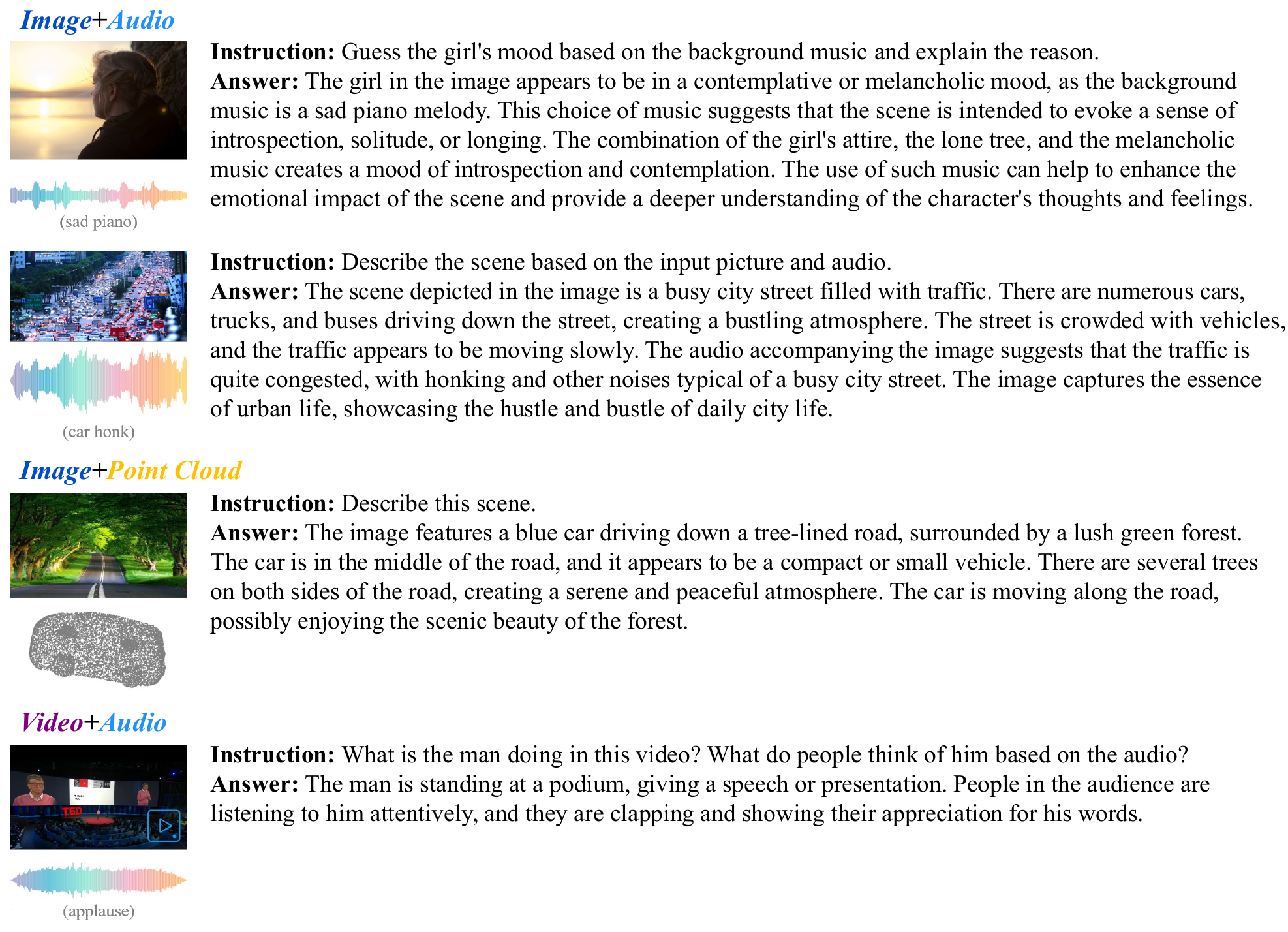}
    \caption{\textbf{Mixture of Modalities with ImageBind-LLM}.}
    \label{fig:mix_modality}
\end{figure*}

\begin{figure*}[htbp]
    \includegraphics[width=\textwidth]{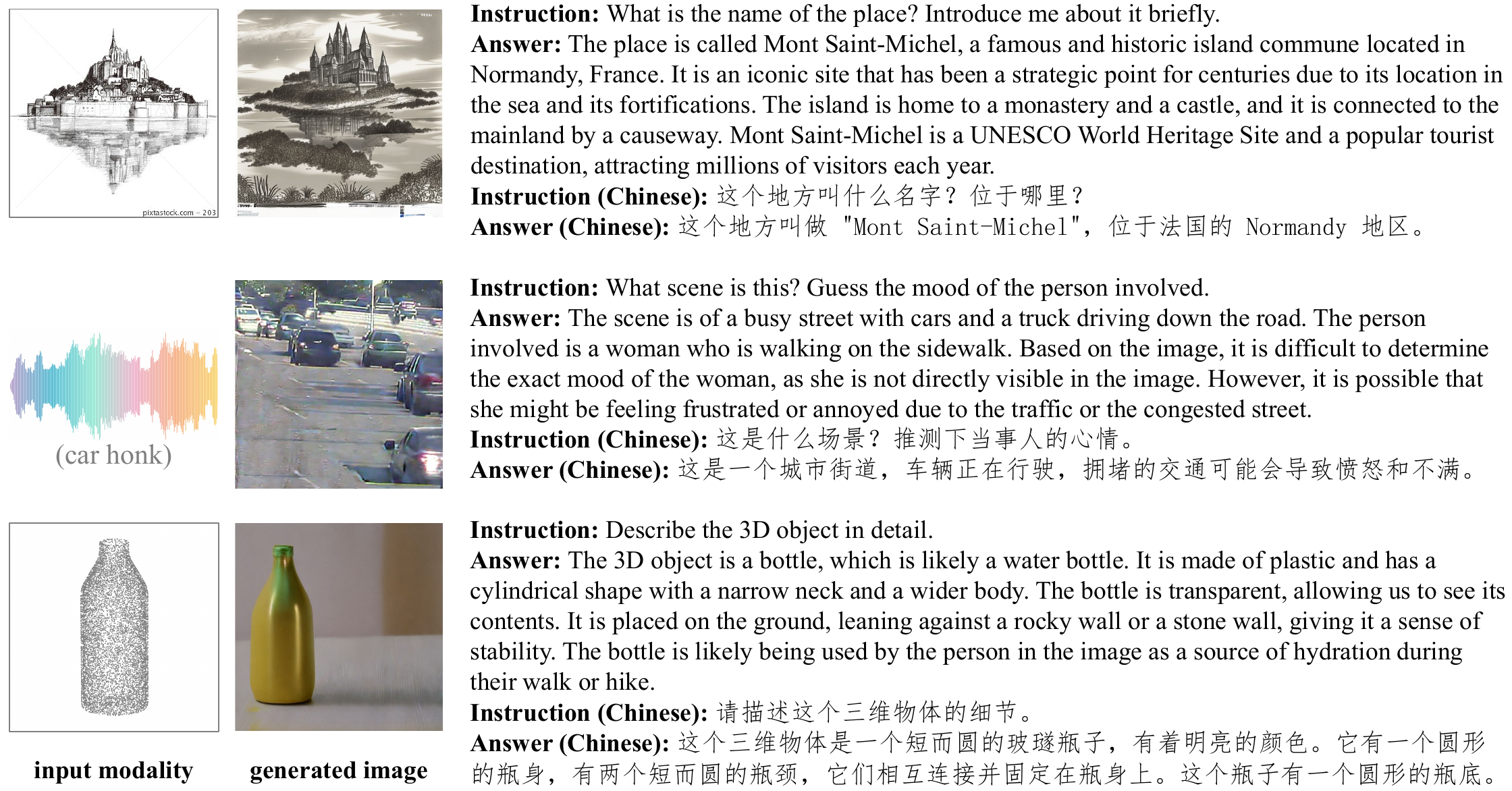}
    \caption{\textbf{Advanced Examples with Bilingual ImageBind-LLM and Integration with Stable Diffusion.}}
    \label{fig:bilingual_and_diffusion}
\end{figure*}

\begin{figure*}[htbp]
    \includegraphics[width=\textwidth]{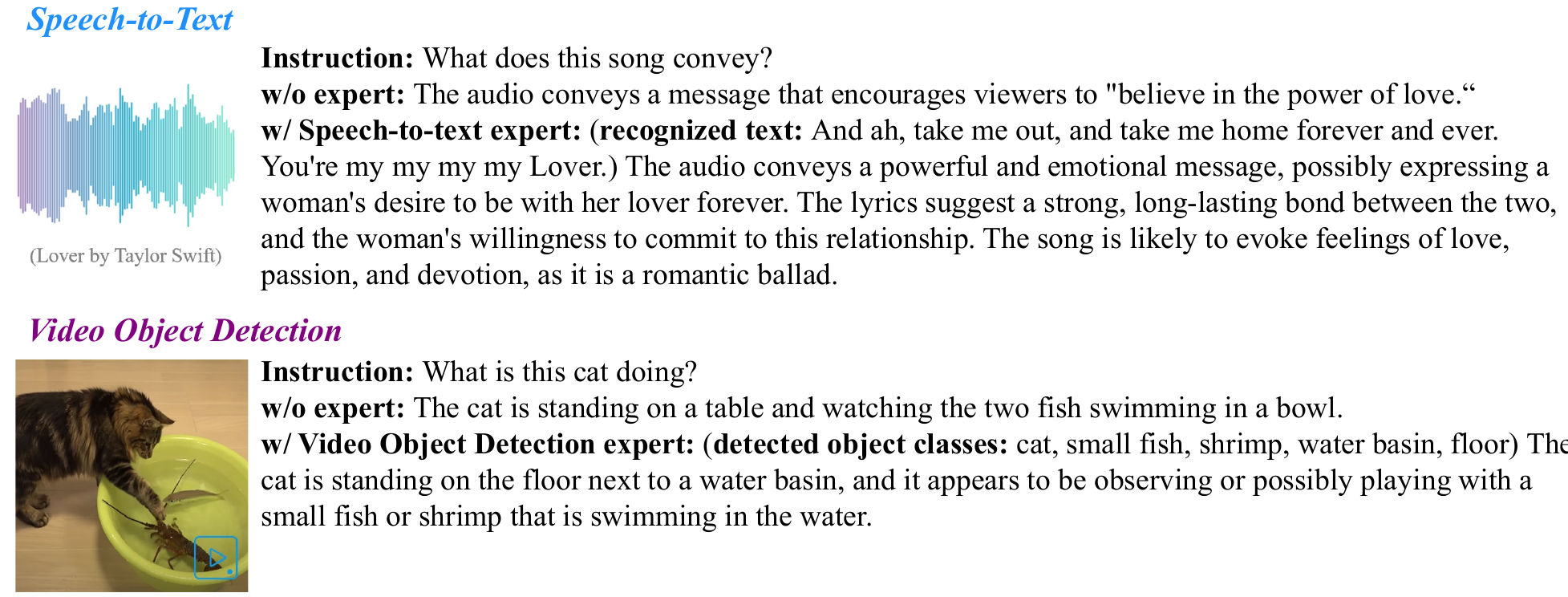}
    \caption{\textbf{Advanced Examples by Integrating ImageBind-LLM with Modality Experts}.}
    \label{fig:domain_expert}
\end{figure*}

\begin{figure*}[htbp]
    \includegraphics[width=\textwidth]{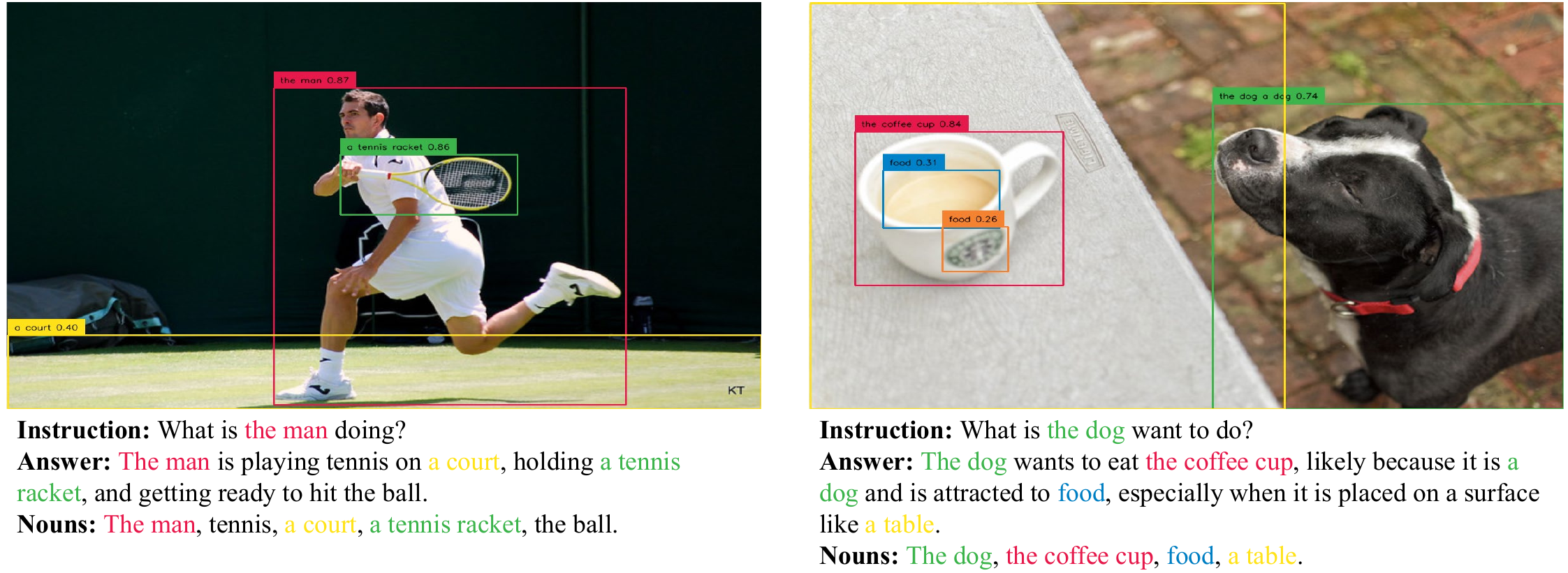}
    \caption{\textbf{Advanced Examples with Object Detection Cascade.}}
    \label{fig:object_cascade}
\end{figure*}

\begin{figure*}[htbp]
  \centering
\includegraphics[width=\textwidth]{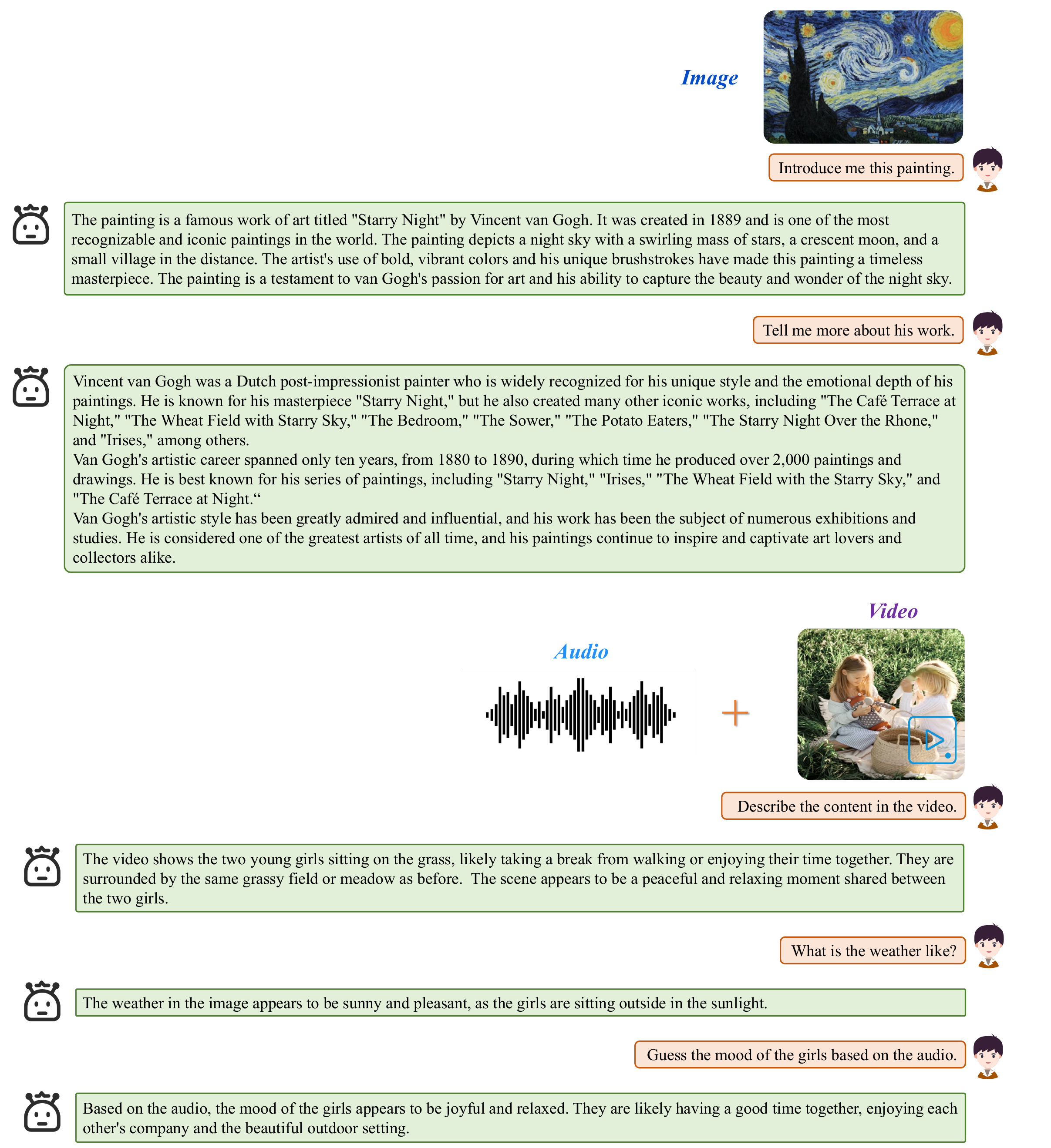}
   \caption{\textbf{Advanced Examples with ImageBind-LLM as Chatbot.}}
    \label{fig:dialog}
\end{figure*}

\begin{figure*}[htbp]
  \centering
\includegraphics[width=\textwidth]{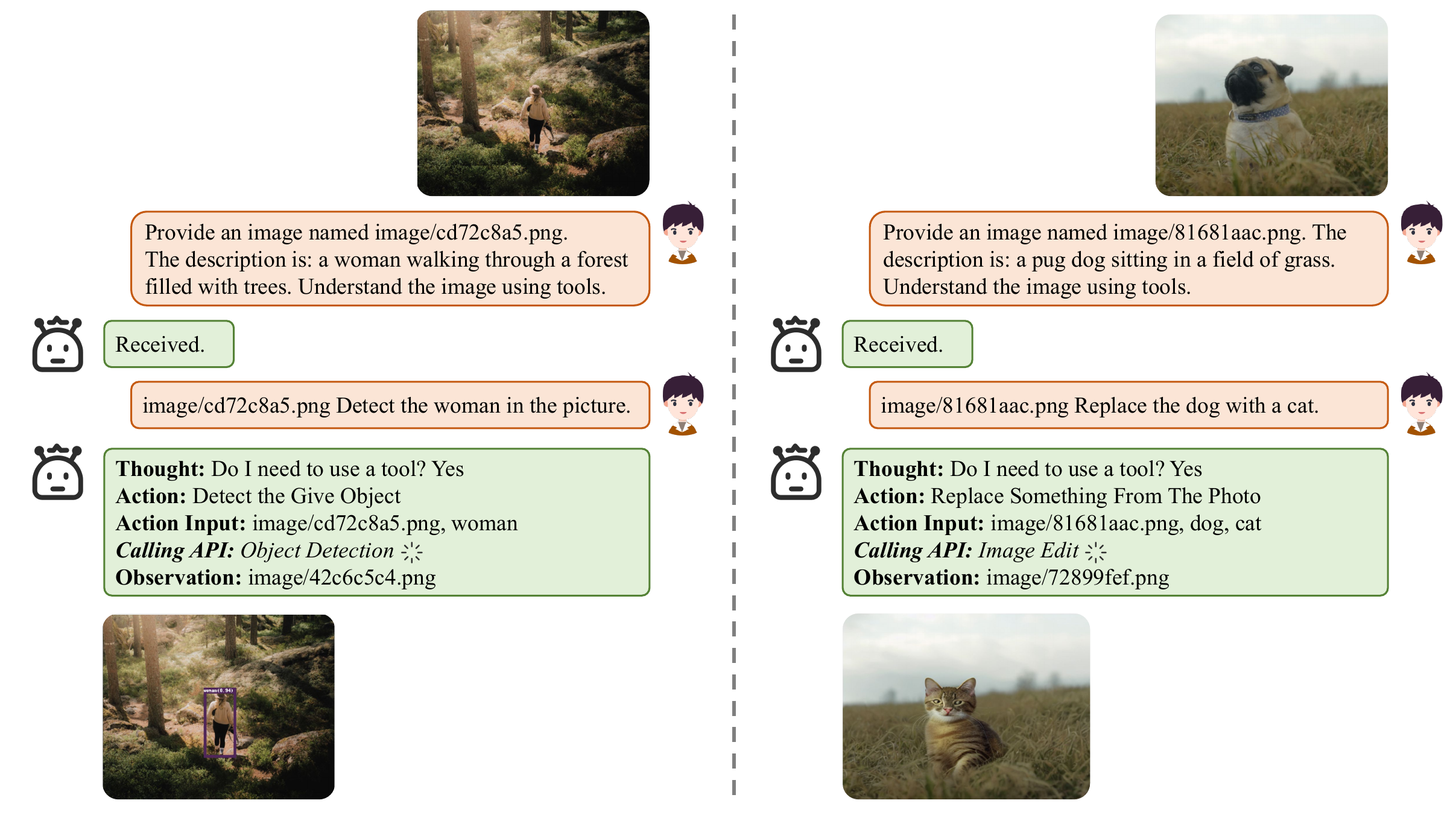}
   \caption{\textbf{Advanced Examples with ImageBind-LLM for API Control.}}
    \label{fig:api}
\end{figure*}

\begin{figure*}[htbp]
    \includegraphics[width=\textwidth]{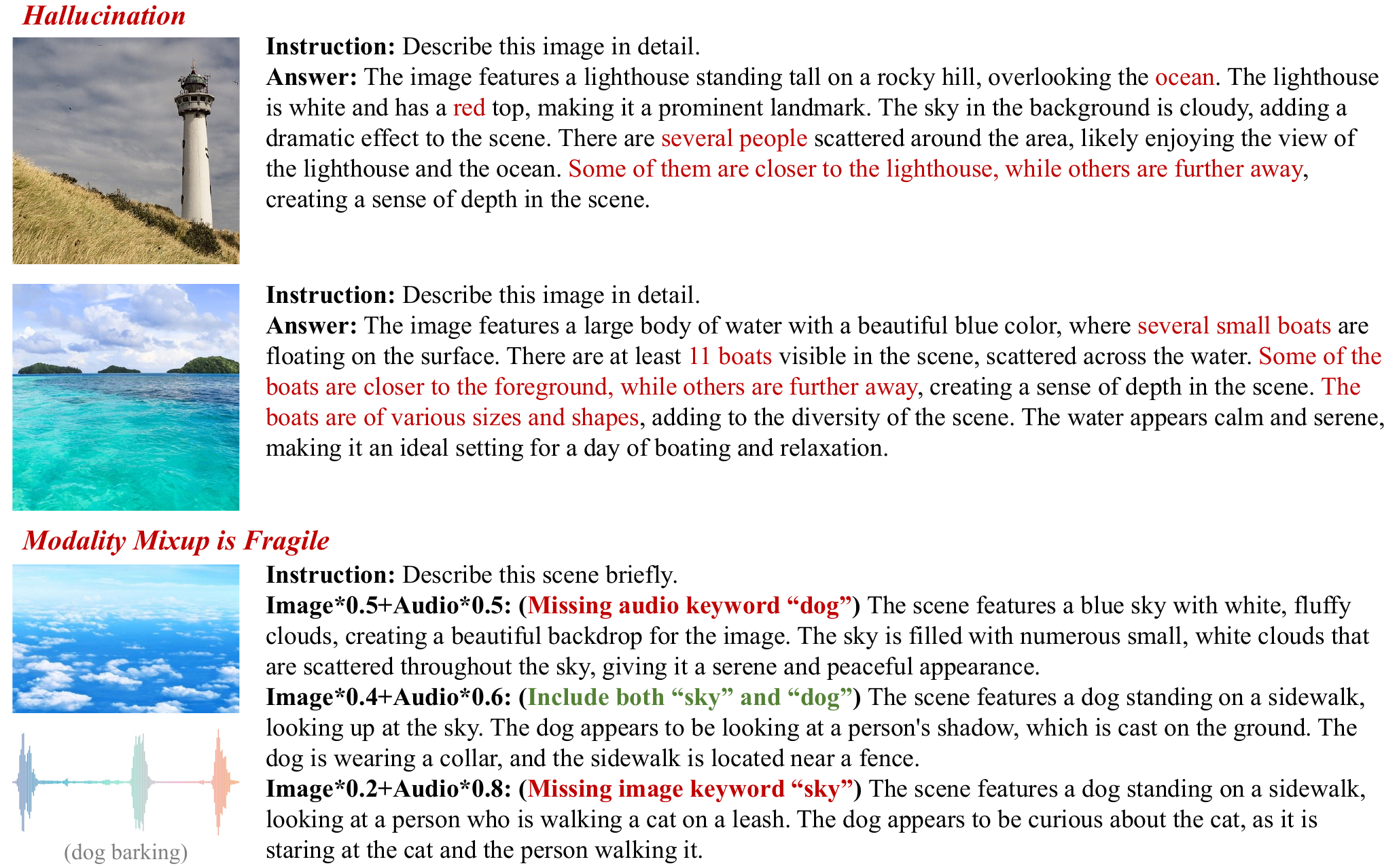}
    \caption{\textbf{Failure Cases.}}
    \label{fig:failure_cases}
\end{figure*}

\subsection{Qualitative Analysis}
In this section, we will give qualitative examples and analysis to help understand how ImageBind-LLM works, and where its multi-modal instruction capabilities come from.

\subsubsection{Multi-modality Understanding}

\noindent \textbf{Multi-modality to Text Alignment.} The vision-language pre-training stage is essential for incorporating multi-modal information into LLMs. In Fig.~\ref{fig:image_caption}, we give some multi-modality captioning results using the pre-trained ImageBind-LLM. As we can see, ImageBind-LLM can generate modality-dependent outputs for image, audio, video and point cloud. Since ImageBind-LLM is pre-trained with image-text pairs, it can give a short and accurate description of the image. Thanks to the binding property of ImageBind, ImageBind-LLM are able to connect other modalities with LLMs without any retraining. Although the pre-trained ImageBind-LLM cannot generate long and detailed description about the input, we believe that the basic ability acquired at this stage is very important for subsequent instruction tuning.

\noindent \textbf{Multi-modality Instruction Tuning.} By fine-tuning ImageBind-LLM on the mixture of language instruction data and visual instruction data, we equip ImageBind-LLM with both language and visual instruction-following abilities. More importantly, we can enter free-form instructions for modals other than images. Taking Fig.~\ref{fig:instruction_demo} as an example, ImageBind-LLM can not only answer questions about images, but also follow instructions of other modalities. We can ask the model to write a poem based on an audio (4th item in Fig.~\ref{fig:instruction_demo}), introduce the content of a movie (5th item in Fig.~\ref{fig:instruction_demo}) and even describe a 3D object (7-8th item in Fig.~\ref{fig:instruction_demo}).

\subsubsection{Mixture of Modalities}
Since ImageBind-LLM unifies different modalities in one feature space, we can mix different modalities as the final input. In detail, we first employ modality encoders to extract modality tokens. We then mix the modality tokens by a set of coefficients. Finally, mixed tokens are fed into LLMs for multi-modality reasoning and understanding. For example, mixing images and audios allows the model to better understand the current scene (1-2th item in Fig.~\ref{fig:api}); The combination of image and point cloud enrich the content of the picture (3-th item in Fig.~\ref{fig:mix_modality}); Using video frames and corresponding audio as input facilitates video understanding.




\subsection{Advanced Applications}

\noindent \textbf{Bilingual ImageBind-LLM.} By joint fine-tuning the model on both English and Chinese instruction data, we turn ImageBind-LLM into a bilingual model. As shown in Fig.~\ref{fig:bilingual_and_diffusion}, ImageBind-LLM can generate accurate Chinese response for a Chinese instruction.  However, since our Chinese visual instruction capacity is emerged in a zero-shot manner, it usually fails to generate long and detailed Chinese responses. We believe that through a stronger bilingual LLM and collecting Chinese visual instruction data, its Chinese instruction capacity can be improved in the future.

\noindent \textbf{Integration with Stable Diffusion.} With LLMs, we realize any-to-language decoding. Similarly, we employ Stable Diffusion for any-to image decoding. As shown in Fig.~\ref{fig:bilingual_and_diffusion}, ImageBind-LLM generates both language and image responses for a give input (\emph{e.g.}, image, audio and point cloud). Compared with language response, the generated image carries more details about the input. For example, an image generated from a car horn can clearly depict the scene and its main elements.

\noindent \textbf{Integration with Modality Experts.} ImageBind-LLM is a general multi-modality LLM, but it still lacks fine-grained domain-specific knowledge. Therefore, integrating with domain experts is a feasible way to improve ImageBind-LLM's multi-modality understanding capacity. In Fig.~\ref{fig:domain_expert}, we use the output of an speech-to-text expert as context to enhance ImageBind-LLM's audio understanding capacity. We also show that class labels extracted by a video object detector can help the model catch details of a video clip.

\noindent \textbf{ImageBind-LLM with Object Detection Cascade.} By cascading ImageBind-LLM with object detectors, we can associate text response with objects in the image, which helps improve the reliability of multi-modal LLMs. As shown in Fig.~\ref{fig:object_cascade}, we can evaluate the reliability of ImageBind-LLM through detection results. At the same time, ImageBind-LLM makes object detectors also has the ability to reasoning.

\noindent \textbf{ImageBind-LLM as Chatbot.} Leveraging multi-turn conversation data for training, ImageBind-LLM showcases impressive visual understanding abilities and multimodal conversation skills. As shown in Fig.~\ref{fig:dialog}, ImageBind-LLM excels in these capabilities. Furthermore, it exhibits the capacity to comprehend and reason with multimodal content in response to user queries, making it a viable candidate for a multi-modality chatbot.

\noindent \textbf{ImageBind-LLM for API control.} By incorporating tool-usage knowledge into LLMs, ImageBind-LLM can effectively learn to invoke API controls, enabling it to tackle various visual tasks. As illustrated in Fig.~\ref{fig:api}, the finetuned ImageBind-LLM demonstrates its proficiency in accomplishing diverse visual tasks, including but not limited to object detection and image editing, by effectively employing different APIs. These results highlight the potential of the proposed ImageBind-LLM in the context of multi-modal tool usage.

\subsection{Failure Cases}

Although ImageBind-LLM can handle multi-modality inputs simultaneously, it is not perfect. As discussed in the previous quantitative analysis, ImageBind-LLM is weak compared to other VLMs. Firstly, ImageBind-LLM often suffers from hallucination issues for descriptive instructions. As shown in Fig.~\ref{fig:failure_cases}, ImageBind-LLM tends to describe objects not shown in the image. There maybe two possible reasons: (1) ImageBind-LLM only injects one global visual token into LLMs, which is much smaller than other models (10 for LLaMA-Adapter, 32 for MiniGPT4 and 256 for LLaVA). Therefore, LLM cannot get enough visual information in the self-Attention layers. (2) The quality of instruction tuning data is not high enough. For example, the visual instruction data from LLaVA is all generated by vision experts and GPT4, lacking human checks and corrections. Therefore, we will build a human-verified high-quality dataset in the future. Secondly, modality mixup is fragile when the two modalities represent different concepts. Fig.~\ref{fig:failure_cases} gives an example of mixing a "sky" image and a "dog barking" audio. We can see that ImageBind-LLM is sensitive to the modality mixing ratio.

\section{Conclusion}
In this work, we propose to tune LLMs into multi-modality instruction models with the help of ImageBind, named ImageBind-LLM. In contrast to prior language instruction models and image instruction models, ImageBind-LLM unifies image, audio, 3D point clouds and video into one multi-modality LLM. We achieves this by simply aligning ImageBind's visual encoder with an LLM via a learnable bind network. Thanks to the binding property of ImageBind, we can directly feed multi-modality inputs into ImageBind-LLM for inference without any training. We also propose a training-free image cache model to mitigate the modality discrepancy between training and inference. We evaluate ImageBind-LLM across 27 traditional vision-language datasets and a new multimodal LLM benchmark MME, where ImageBind-LLM achieves comparable performance with recent mutlimodal LLMs. We also give extensive qualitative analysis to demonstrate ImageInd-LLM's multi-modality understanding capacity. In the future, we plan to enhance ImageBind-LLM by increasing the number of multi-modality tokens. Besides, integrating more modalities into ImageBind-LLM is also a promising approach to enhance its multimodal capacity.




\bibliographystyle{sn-mathphys.bst}
\bibliography{sn-bibliography}

\end{document}